\documentclass[draft]{agujournal}
\usepackage{amsmath, amsthm, amssymb, amsfonts}
\begin{document}

\title{Gamma-ray Showers Observed at Ground Level in Coincidence With Downward Lightning Leaders}

%%%%%%%%%%%%%%%%%%%%%%%%%%%%%%%%%%%%%%%%

\authors{
R.U.~Abbasi\affil{1}, 
T.Abu-Zayyad \affil{1}, 
M.~Allen \affil{1}, 
E.~Barcikowski \affil{1}, 
J.W.~Belz\affil{1}, 
D.R.~Bergman\affil{1}, 
S.A.~Blake\affil{1}, 
M.~Byrne\affil{1},
R.~Cady\affil{1}, 
B.G.~Cheon\affil{4},
J.~Chiba\affil{5}, 
M.~Chikawa\affil{6}, 
T.~Fujii\affil{7}, 
M.~Fukushima\affil{7,8},
G.~Furlich\affil{1}, 
T.~Goto\affil{9}, 
W.~Hanlon\affil{1}, 
Y.~Hayashi\affil{9}, 
N.~Hayashida\affil{10}, 
K.~Hibino\affil{10}, 
K.~Honda\affil{11}, 
D.~Ikeda\affil{7}, 
N.~Inoue\affil{2}, 
T.~Ishii\affil{11}, 
H.~Ito\affil{12}, 
D.~Ivanov\affil{1},
S.~Jeong\affil{13},
C.C.H.~Jui\affil{1}, 
K.~Kadota\affil{14}, 
F.~Kakimoto\affil{3}, 
O.~Kalashev\affil{15},
K.~Kasahara\affil{16}, 
H.~Kawai\affil{17}, 
S.~Kawakami\affil{9}, 
K.~Kawata\affil{7}, 
E.~Kido\affil{7}, 
H.B.~Kim\affil{4}, 
J.H.~Kim\affil{1}, 
J.H.~Kim\affil{18}, 
S.S.~Kishigami\affil{9}, 
P.R.~Krehbiel\affil{19},
V.~Kuzmin\affil{15}, 
Y.J.~Kwon\affil{20}, 
J.~Lan\affil{1},
R.~LeVon\affil{1}, 
J.P.~Lundquist\affil{1}, 
K.~Machida\affil{11}, 
K.~Martens\affil{8}, 
T.~Matuyama\affil{9}, 
J.N.~Matthews\affil{1}, 
M.~Minamino\affil{9}, 
K.~Mukai\affil{11}, 
I.~Myers\affil{1}, 
S.~Nagataki\affil{12}, 
R.~Nakamura\affil{27}, 
T.~Nakamura\affil{22}, 
T.~Nonaka\affil{7}, 
S.~Ogio\affil{9}, 
M.~Ohnishi\affil{7}, 
H.~Ohoka\affil{7}, 
K.~Oki\affil{7}, 
T.~Okuda\affil{23}, 
M.~Ono\affil{24}, 
R.~Onogi\affil{9}, 
A.~Oshima\affil{25}, 
S.~Ozawa\affil{16}, 
I.H.~Park\affil{13}, 
M.S.~Pshirkov\affil{14,26}, 
J.~Remington\affil{1},
W.~Rison\affil{19},
D.~Rodeheffer\affil{19},
D.C.~Rodriguez\affil{1}, 
G.~Rubtsov\affil{15}, 
D.~Ryu\affil{18}, 
H.~Sagawa\affil{7}, 
K.~Saito\affil{7}, 
N.~Sakaki\affil{7}, 
N.~Sakurai\affil{9}, 
T.~Seki\affil{27}, 
K.~Sekino\affil{7}, 
P.D.~Shah\affil{1}, 
F.~Shibata\affil{11}, 
T.~Shibata\affil{7}, 
H.~Shimodaira\affil{7}, 
B.K.~Shin\affil{9}, 
H.S.~Shin\affil{7}, 
J.D.~Smith\affil{1}, 
P.~Sokolsky\affil{1},
R.W.~Springer\affil{1}, 
B.T.~Stokes\affil{1}, 
T.A.~Stroman\affil{1}, 
H.~Takai\affil{29}, 
M.~Takeda\affil{7}, 
R.~Takeishi\affil{7}, 
A.~Taketa\affil{30}, 
M.~Takita\affil{7}, 
Y.~Tameda\affil{31},
H.~Tanaka\affil{9}, 
K.~Tanaka\affil{32}, 
M.~Tanaka\affil{21}, 
R.J.~Thomas\affil{19},
S.B.~Thomas\affil{1}, 
G.B.~Thomson\affil{1}, 
P.~Tinyakov\affil{14,32}, 
I.~Tkachev\affil{15}, 
H.~Tokuno\affil{3}, 
T.~Tomida\affil{27}, 
S.~Troitsky\affil{15}, 
Y.~Tsunesada\affil{9}, 
Y.~Uchihori\affil{34}, 
S.~Udo\affil{10}, 
F.~Urban\affil{33}, 
G.~Vasiloff\affil{1}, 
T.~Wong\affil{1}, 
M. Yamamoto\affil{27}, 
R.~Yamane\affil{9}, 
H.~Yamaoka\affil{21}, 
K.~Yamazaki\affil{30}, 
J.~Yang\affil{35}, 
K.~Yashiro\affil{5}, 
Y.~Yoneda\affil{9}, 
S.~Yoshida\affil{17}, 
H.~Yoshii\affil{36}, 
Z.~Zundel\affil{1} 
} % end of authors

\affiliation{1}{High Energy Astrophysics Institute and Department of Physics and Astronomy, University of Utah, Salt Lake City, Utah, USA}
\affiliation{2}{The Graduate School of Science and Engineering, Saitama University, Saitama, Saitama, Japan}
\affiliation{3}{Graduate School of Science and Engineering, Tokyo Institute of Technology, Meguro, Tokyo, Japan}
\affiliation{4}{Department of Physics and The Research Institute of Natural Science, Hanyang University, Seongdong-gu, Seoul, Korea}
\affiliation{5}{Department of Physics, Tokyo University of Science, Noda, Chiba, Japan}
\affiliation{6}{Department of Physics, Kinki University, Higashi Osaka, Osaka, Japan}
\affiliation{7}{Institute for Cosmic Ray Research, University of Tokyo, Kashiwa, Chiba, Japan}
\affiliation{8}{Kavli Institute for the Physics and Mathematics of the Universe (WPI), Todai Institutes for Advanced Study, the University of Tokyo, Kashiwa, Chiba, Japan}
\affiliation{9}{Graduate School of Science, Osaka City University, Osaka, Osaka, Japan}
\affiliation{10}{Faculty of Engineering, Kanagawa University, Yokohama, Kanagawa, Japan}
\affiliation{11}{Interdisciplinary Graduate School of Medicine and Engineering, University of Yamanashi, Kofu, Yamanashi, Japan}
\affiliation{12}{Astrophysical Big Bang Laboratory, RIKEN, Wako, Saitama, Japan}
\affiliation{13}{Department of Physics, Sungkyunkwan University, Jang-an-gu, Suwon, Korea} 
\affiliation{14}{Department of Physics, Tokyo City University, Setagaya-ku, Tokyo, Japan}
\affiliation{15}{National Nuclear Research University, Moscow Engineering Physics Institute, Moscow, Russia}
\affiliation{16}{Advanced Research Institute for Science and Engineering, Waseda University, Shinjuku-ku, Tokyo, Japan}
\affiliation{17}{Department of Physics, Chiba University, Chiba, Chiba, Japan}
\affiliation{18}{Department of Physics, School of Natural Sciences, Ulsan National Institute of Science and Technology, UNIST-gil, Ulsan, Korea}
\affiliation{19}{Langmuir Laboratory for Atmospheric Research, New Mexico Institute of Mining and Technology, Socorro, New~Mexico 87801, USA.}
\affiliation{20}{Department of Physics, Yonsei University, Seodaemun-gu, Seoul, Korea}
\affiliation{21}{Institute of Particle and Nuclear Studies, KEK, Tsukuba, Ibaraki, Japan}
\affiliation{22}{Faculty of Science, Kochi University, Kochi, Kochi, Japan}
\affiliation{23}{Department of Physical Sciences, Ritsumeikan University, Kusatsu, Shiga, Japan}
\affiliation{24}{Department of Physics, Kyushu University, Fukuoka, Fukuoka, Japan}
\affiliation{25}{Engineering Science Laboratory, Chubu University, Kasugai, Aichi, Japan}
\affiliation{26}{Sternberg Astronomical Institute Moscow M.V.Lomonosov State University, Moscow, Russia}
\affiliation{27}{Department of Computer Science and Engineering, Shinshu University, Nagano, Nagano, Japan}
\affiliation{28}{Department of Physics and Astronomy, Rutgers University - The State University of New Jersey, Piscataway, New Jersey, USA}
\affiliation{29}{Brookhaven National Laboratory, Upton, New York, USA}
\affiliation{30}{Earthquake Research Institute, University of Tokyo, Bunkyo-ku, Tokyo, Japan}
\affiliation{31}{Department of Engineering Science, Faculty of Engineering, Osaka Electro-Communication University, Neyagawa, Osaka,  JAPAN}
\affiliation{32}{Graduate School of Information Sciences, Hiroshima City University, Hiroshima, Hiroshima, Japan}
\affiliation{33}{Service de Physique Th$\acute{\rm e}$orique, Universit$\acute{\rm e}$ Libre de Bruxelles, Brussels, Belgium}
\affiliation{34}{National Institute of Radiological Science, Chiba, Chiba, Japan}
\affiliation{35}{Department of Physics and Institute for the Early Universe, Ewha Womans University, Seodaaemun-gu, Seoul, Korea}
\affiliation{36}{Department of Physics, Ehime University, Matsuyama, Ehime, Japan}

%%%%%%%%%%%%%%%%%%%%%%%%%%%%%%%%%%%%%%%%
%\correspondingauthor{C.~Author}{email@address.edu}

\begin{keypoints}

\item Gamma-ray showers have been detected in a surface scintillator array coincident with lightning observed by a lightning mapping array or $\Delta$E antenna. 
\item The showers were produced less than 4-5 kilometers above ground in the first 1-2~milliseconds of downward negative breakdown during cloud-to-ground flashes.

\item  The source durations are better resolved than for satellite observations and are consistent with being produced by stepping of the initial leader breakdown.

\end{keypoints}

\begin{abstract}

Bursts of gamma ray showers have been observed in coincidence with downward propagating negative leaders in lightning flashes by the Telescope Array Surface Detector (TASD). The TASD is a 700~square kilometer cosmic ray observatory located in southwestern Utah, U.S.A. In data collected between 2014 and 2016, correlated observations showing the structure and temporal development of three shower-producing flashes were obtained with a 3D lightning mapping array, and electric field change measurements were obtained for an additional seven flashes, in both cases co-located with the TASD.  National Lightning Detection Network (NLDN) information was also used throughout. The showers arrived in a sequence of 2--5 short-duration ($\le$10~$\mu$s) bursts over time intervals of several hundred microseconds, and originated at an altitude of $\simeq$3--5 kilometers above ground level during the first 1--2 ms of downward negative leader breakdown at the beginning of cloud-to-ground lightning flashes. The shower footprints, associated waveforms and the effect of atmospheric propagation indicate that the showers consist primarily of downward-beamed gamma radiation. 
%This has been supported by GEANT simulation studies, which indicate primary source fluxes of $\simeq 10^{14}$ down to $10^{12}$ photons for narrow ($16^{\circ}$) angular beams, and an order of magnitude larger for wider beaming.  
This has been supported by GEANT simulation studies, which indicate primary source fluxes of $\simeq$$10^{12}$--$10^{14}$ photons for $16^{\circ}$ half-angle beams.  We conclude that the showers are terrestrial gamma-ray flashes (TGFs), similar to those observed by satellites, but that the ground-based observations are more representative of the temporal source activity and are also more sensitive than satellite observations, which detect only the most powerful TGFs.

\end{abstract}

%%%%%%%%%%%%%%%%%%%%%%%%%%%%%%%%%%%%%%%%%%%%%%%
%input{introduction_grl.tex }
\section{Introduction}

Terrestrial gamma-ray flashes (TGFs) are bursts of gamma-rays initiated in the Earth's atmosphere, first reported in 1994 from  data collected with the Burst and Transient Source Experiment (BATSE) on the Compton Gamma-Ray Observatory satellite~\citep{fishman1994,BATSE1994}.  Since then, a number of observations have shown that satellite-detected TGFs are produced by lightning flashes. In particular, the observations indicate that the TGFs occur within the first few milliseconds of upward intracloud (IC) flashes~\citep{Stanley2006,shao2010,lu2010,cummer2011,cummer2015, lyu2016}. Concurrently, the observations were found to be consistent with  the relativistic runaway electron avalanche (RREA) model~\citep{GRL:GRL20344,dwyer2008,dwyer2012a,dwyer_etal2012}. In normally-electrified storms, intracloud flashes occur between the main mid-level negative and the upper positive charge region. They typically begin with upward-developing negative breakdown (e.g. \cite{shao1996} and \cite{behnke2005}), thus their detection by overhead satellites.

Since their discovery, an important question has been whether TGFs can be detected at ground level. Four gamma-ray observations have been detected by ground experiments: two of the experiments have reported observations of gamma-rays after return strokes, indicating that gamma-rays seen on the ground may originate from a mechanism different from that of the TGF satellite events. \cite{Dwyer2012} at the International Center for Lightning Research and Testing (ICLRT) and \cite{Tran201586} at the Lightning Observatory in Gainesville (LOG) reported observing gamma-rays about 200~$\mu$s after the beginning of the upward return stroke of  natural negative cloud-to-ground (--CG) flashes. The other two, ~\cite{Dwyer2004} and~\cite{hare2016} at the ICLRT reported gamma-ray observations in association with upward positive leaders in rocket-triggered lightning. Both occurred several kilometers Above Ground Level (AGL). More recently, \cite{bowers2017} and \cite{enoto2017} presented ground-based observations of neutron-producing TGFs coincident with return strokes of $-$CG flashes in Japan. In the first case the stroke was triggered by upward positive breakdown from a wind turbine, and in the second case the stroke was produced by apparent natural downward breakdown during a low-altitude winter storm.

The fact that satellite-detected TGFs appear to be produced during upward negative breakdown at the beginning of intracloud discharges suggests that TGFs should also be produced by the downward negative breakdown that occurs at the beginning of --CG flashes. In this paper we present the first observations that this indeed happens. 

A previous report of Telescope Array Surface Detector (TASD) data collected between 2008--2013 showed a strong correlation between bursts of energetic particle showers and NLDN lightning activity~\citep{Abbasi20172565}. Here, we extend those studies  with new data collected simultaneously with local Lightning Mapping Array (LMA) and electrostatic field change ($\Delta E$) measurements as well as with detailed simulation studies. Over a two-year period between 2014 and 2016, a total of ten TGF bursts were identified for which 3-D LMA or $\Delta$E lightning measurements were available. In each case the parent flash was a --CG discharge and the burst occurred within the first or second millisecond of the flash. The combined data from the enhanced set of instruments, along with simulation studies of the TASD response to atmospheric photons, firmly establish that these shower bursts are consistent with downward TGFs.

Previous ground--based $x$-- and gamma--ray observations have primarily utilized NaI detectors with limited areal coverage.  The TASD of the present study utilizes large-area (3~m$^2$) plastic scintillators that are optimized for detecting high-energy charged particles produced in cosmic ray air showers.  While lacking the ability to measure the energy of individual particles, the TASD response time is roughly ten times faster than that of NaI detectors. Moreover, the TASD covers an area hundreds of times larger than other ground-based detectors of lightning-associated events, making it the largest such detector to date.  The addition of an LMA network and $\Delta$E  observations to the TASD has provided us with a unique suite of instruments for studying the TGF phenomena.

%%%%%%%%%%%%%%%%%%%%%%%%%%%%%%%%%%%%%%%%%%%%%%%
%input{detectors_grl.tex}
%\vspace{-1\baselineskip}

\section{Telescope Array, Lightning Mapping Array, and Slow Antenna Detectors}

\subsection{The Telescope Array Surface Detector}
\label{sec:ta}

\begin{figure}
\includegraphics[width=\textwidth]{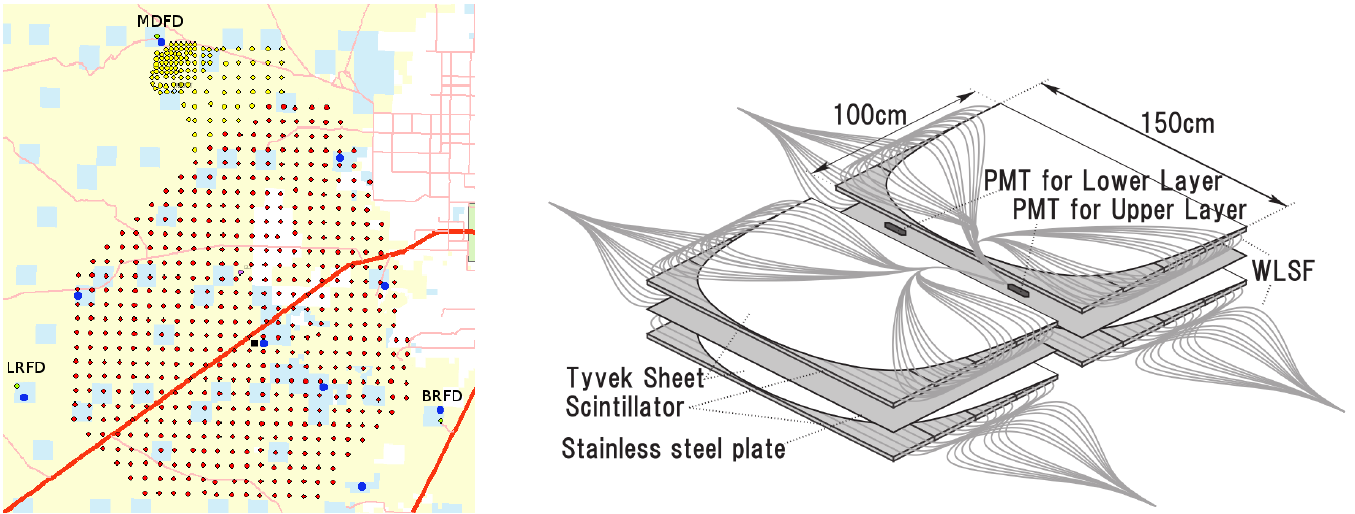}
\caption{{\em Left:} The Telescope Array, consisting of 507 scintillator Surface Detectors (SDs) on a 1.2~km grid over a 700~km$^2$ area (red dots), and three Fluorescence Detectors (MDFD, LRFD, BRFD). Nine LMA stations (blue dots) are located within and around the array, with the slow E sensor (SA) close to the central LMA station (black square).  {\em Right:} Schematic sketch of the upper and lower 1~cm thick plastic scintillator layers inside the scintillator box, the 1~mm stainless steel plate, the 104 wavelength-shifting (WLS) fibers and the photomultiplier tubes (PMTs). These items are enclosed in a stainless steel box, ~1.5~mm thick on top and 1.2~mm thick on the bottom.} 
%\vspace{-3\baselineskip}
\label{fig:TAmap} 
\end{figure}

The Telescope Array (TA) is located in the southwestern desert of the State of Utah, and was commissioned with the primary goal of detecting Ultra High Energy Cosmic Rays (UHECRs).  It is composed of a 700 km$^2$ Surface Detector array (SD), overlooked by three Fluorescence Detector (FD) sites~\citep{AbuZayyad:2012kk} (Figure~\ref{fig:TAmap}, left).  The FD, which operates on clear moonless nights (approximately 10\% duty cycle) provides a measurement of the longitudinal profile of the Extensive Air Shower (EAS) induced by the primary UHECR, as well as a calorimetric estimate of the EAS energy. The SD part of the detector, with approximately 100\% duty cycle, provides shower footprint information including core location, lateral density profile, and timing, which are used to reconstruct shower geometry and energy.

The TASD is currently comprised of 507 scintillator detectors on a 1.2~km square grid. Each detector unit consists of upper and lower scintillator planes, each plane is 3~m$^2$ in area by 1~cm thick (Figure~\ref{fig:TAmap}, right). The upper and lower planes are separated by a 1~mm thick steel plate, and are read out by individual photomultiplier tubes (PMTs) which are coupled to the scintillator via an array of wavelength-shifting fibers. The scintillator, fibers and photomultipliers are contained in a light-tight and electrically grounded stainless steel box (1.5~mm thick on top and 1.2~mm thick on the bottom) under an additional 1.2~mm iron roof providing protection from extreme temperature variations~\citep{AbuZayyad:2012kk}. 

The output signals from the PMTs are digitized locally by a 12~bit Fast Analog-to-Digital Converter (FADC) with a 50 MHz sampling rate~\citep{nonaka:2009icrc}. Each detector unit also has a 1~m$^2$ solar panel, a stainless steel box placed under the solar panel housing the battery and the front-end electronics, and a 2.4~GHz wireless LAN modem communication antenna transmitting data to communication towers.

The TASD is designed to detect the charged components (primarily electrons, positrons, and muons) of the EAS with a timing precision of 20~ns. An event trigger is recorded when three adjacent SDs observe a time-integrated signal greater than 3~Vertical Equivalent Muon (VEM) within 8~$\mu$s. The VEM is a unit of energy deposit, equivalent to the energy deposited in a single TASD scintillator plane by a vertical (hence perpendicular to the plane) relativistic muon. In more conventional units a VEM is about 2~MeV per scintillator plane, roughly 30~ADC counts above background with a pulse 100~ns FWHM. The abundance of penetrating cosmic-ray induced muons in the Earth's atmosphere makes the VEM a convenient standard for scintillation detectors. 

When a trigger occurs, the signals from all the SDs within $\pm$~32~$\mu$s detecting an integrated amplitude greater than 0.3~VEM are also recorded. The TASD typically triggers on cosmic ray air showers about once every hundred seconds. The trigger efficiency of UHECRs with zenith angle less than 45$^\circ$ and energy greater than 10~EeV is approximately 100\%, with a corresponding aperture of 1,100~km$^2$sr~\citep{AbuZayyad:2012kk}.

After a trigger is established and the event recorded, the arrival timing and lateral distribution of the shower particles are used to reconstruct the shower's arrival direction and energy offline. While the TASD is designed to provide good timing information for high-energy charged particles produced in cosmic-ray induced EAS, measuring energy for individual particles is not possible. Rather, the energy of the EAS (and hence the primary cosmic ray) is estimated by counting the effective number of VEM in the individual scintillator as a function of the lateral distance from the shower core and the zenith angle of the shower. Shower energy can be estimated by comparing the lateral distribution with the predictions of high-energy hadronic models, and then by scaling to the calorimetric FD measurement~\citep{AbuZayyad:2012ru}.

\begin{figure}
\begin{center}
\includegraphics[width=\textwidth]{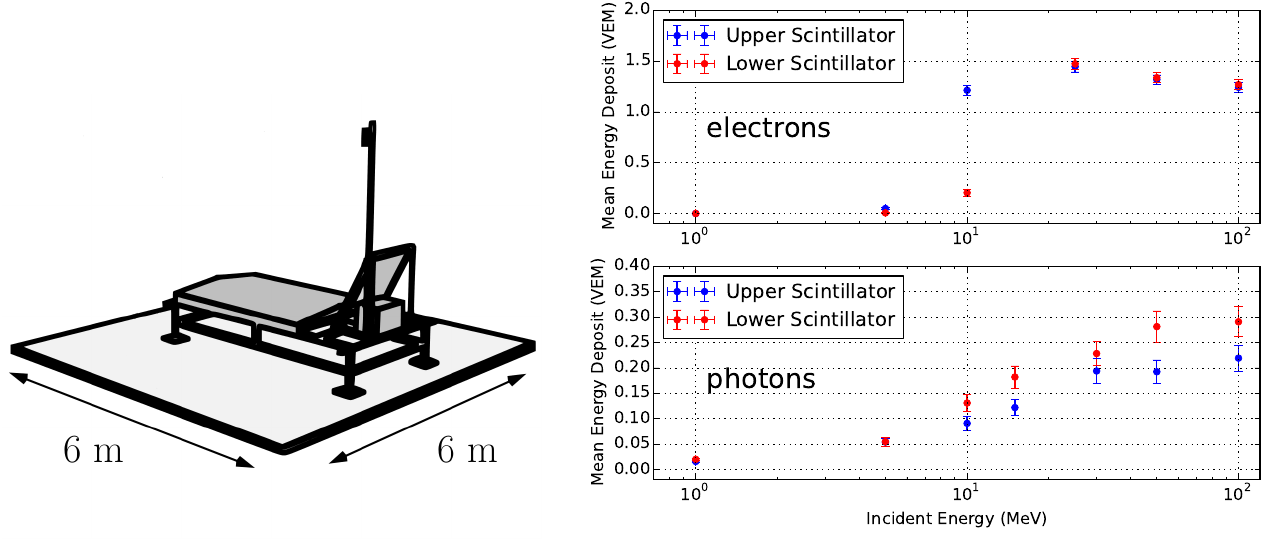}
\caption{{\em Left}: Illustration of the detector model which was simulated in GEANT4~\citep{Agostinelli:2002hh,Allison:2006ve}. The full support structure --- including antenna, solar panel, and electronics box --- was simulated, in addition to the active scintillator planes. Particles were thrown from a height just above the detector, hence the simulation does not include the effects of propagation through the atmosphere. See further description in text. Figure taken from~\cite{ivanov:2012thesis}. {\em Right}: Results of GEANT4 simulation of mean TASD energy deposit versus incident energy for single electrons (top) and photons (bottom) which hit the detector unit.}
\label{fig:depvalt}
\end{center}
\end{figure}

The TASD is an inefficient detector of $x$-ray and gamma radiation, relying on the production of high-energy electrons through the Compton scattering mechanism in either the thin scintillator, steel housing, or air above the detector units. In order to understand the overall response of the detector, we performed a detailed GEANT4~\citep{Agostinelli:2002hh,Allison:2006ve} simulation of the individual scintillator response~\citep{ivanov:2012thesis}. The detector steel and scintillators were included in the simulation, along with supporting structure, as was the earth under the detector which could contribute to signal via backscatter. See Figure~\ref{fig:depvalt}.

From a random starting point at an elevation just above the detector, electrons and photons are initially pointed towards a random coordinate within a $6 \times 6$~m$^2$ area, intentionally larger than the detector itself in order to take into account backscattering effects. Energy deposited in the upper and lower planes of the scintillator is recorded as a function of incident particle energy for a large number of incident primary particles. Mean energy deposit is recorded in the upper and lower scintillator planes as a function of incident particle energy.

As shown in Figure~\ref{fig:depvalt}, electrons above approximately 10~MeV (20~MeV) are expected to deposit the energy equivalent of 1~VEM in the upper (lower) scintillator. Below this, the total energy deposited by electrons falls off rapidly; below 1~MeV there is no detectable energy deposit as the electrons fail to penetrate a significant depth into the scintillator. 

High-energy photons on average will deposit about 20\% (30\%) of a VEM in the upper (lower) scintillator. The majority of photons will not interact in the detector, those that do will primarily create Compton recoil electrons with kinetic energies at or below the photon energy level (see Supporting Figure S9). These electrons can then deposit energy in the scintillator, though the amount deposited in each plane will depend on where the Compton scatter occurs. 

\subsection{The Lightning Mapping Array and Slow Antenna}

The Lightning Mapping Array (LMA) was developed by the Langmuir
Laboratory group at New Mexico Tech
~\citep{GRL:GRL12477,JGRD:JGRD11263}. The LMA produces detailed 3-D
images of the VHF radiation produced by lightning inside storms. A
nine-station network of sensors was deployed in 2013 over the
700~km$^2$ area covered by the TA detector (Figure~\ref{fig:TAmap},
left). It detects the peak arrival time of impulsive radio emissions
in a locally unused TV channel, in this case U.S. Channel 3
(60-66~MHz). In radio-quiet areas such as the southwestern Utah
desert, the LMA detects VHF emissions with a time accuracy of 35~ns
rms over a wide ($>$70~dB) dynamic range, from $\le$10~mW to more than 100~kW peak source power. Cell data modems connect each station into the internet, allowing decimated data to be processed in real time and posted on the web, and for monitoring station operation.

During 2014, a slow antenna (SA) recorded electric field changes of the lightning discharges ~\citep{krehbiel1979}. The SA was located in the center of the TASD and continuously recorded 10 kHz, 24-bit sampled data from a downward-looking flat plate sensor with a time accuracy of 1 $\mu$s. The data were stored locally on 256~GB SD cards, and accurately measured electric field changes over the range of 10~mV/m to 10~kV/m with a decay time constant of 10~seconds.

%%%%%%%%%%%%%%%%%%%%%%%%%%%%%%%%%%%%%%%%%%%%%%%
%\input{observations_grl.tex}
\section{Observations}

Following~\cite{Abbasi20172565}, we searched for candidate lightning events in the TASD dataset by identifying instances in which ``bursts'' of consecutive TASD triggers were recorded in 1~ms time intervals. Since the TASD mean trigger rate from cosmic ray events is less than 0.01~Hz, it is extremely unlikely that such bursts would be caused by the accidental coincidence of high-energy cosmic rays. 

Lightning flashes that produce trigger bursts are rare. There are typically about 750~NLDN-recorded flashes (intracloud and cloud-to-ground) per year over the 700~km$^2$ TASD array. In eight years of TA operation, we have found 20~bursts including the 10 reported in the present paper and 10 reported in~\citep{Abbasi20172565}, which appear to be similar in nature. Thus fewer than one percent of NLDN flashes recorded over the TASD are accompanied by identifiable gamma bursts.

\begin{figure}[h]
\includegraphics[width=\textwidth]{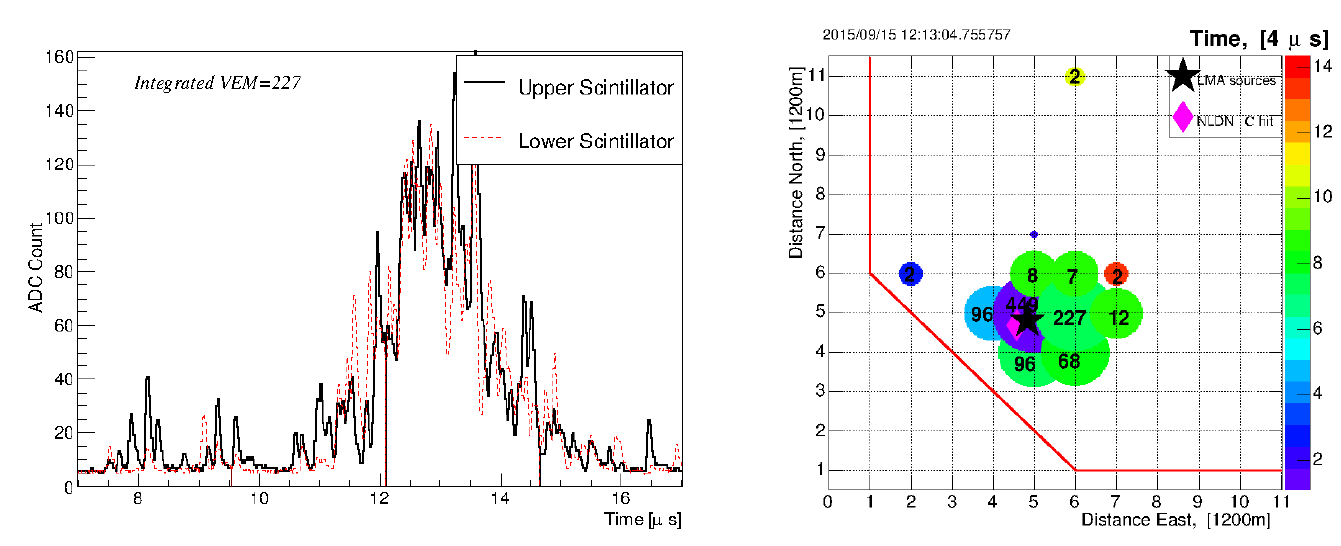}
\caption{{\em Left:} Upper and lower scintillator waveforms in a
  single surface detector unit, for the second trigger in the
  LMA-correlated  energetic radiation burst observed at 12:13:04 on 15 Sept.\ 2015 (FL01, see also Figure~\ref{fig:WFS} and Figure~\ref{fig:new_fig3}a and b).  {\em Right:} Footprint of TASD hits for all detectors units involved in the second trigger of the burst, with the numbers indicating the Vertical Equivalent Muon (VEM) counts (see text), and the color indicating the relative arrival times.  Initial LMA and NLDN events are indicated by stars and diamonds respectively. The red line indicates the southwestern boundary of the TASD array.} 
\label{fig:2figsB} 
\end{figure}

Figure~\ref{fig:2figsB} shows an example of a typical SD waveform and the corresponding TASD footprint from a trigger burst event. As in the standard TASD analysis for cosmic rays, the integrated area under the photomultiplier waveform, relative to that expected for a single minimum-ionizing muon, gives the Vertical Equivalent Muon (VEM) count by which each SD trigger is characterized. One VEM corresponds to $\sim 2$~MeV of energy deposited in the 1~cm thick scintillator. The footprint plot of Figure~\ref{fig:2figsB} shows the VEM counts for each participating surface detector, with area proportional to the logarithm of the total energy deposited.

\begin{figure}[h]
\includegraphics[width=\textwidth]{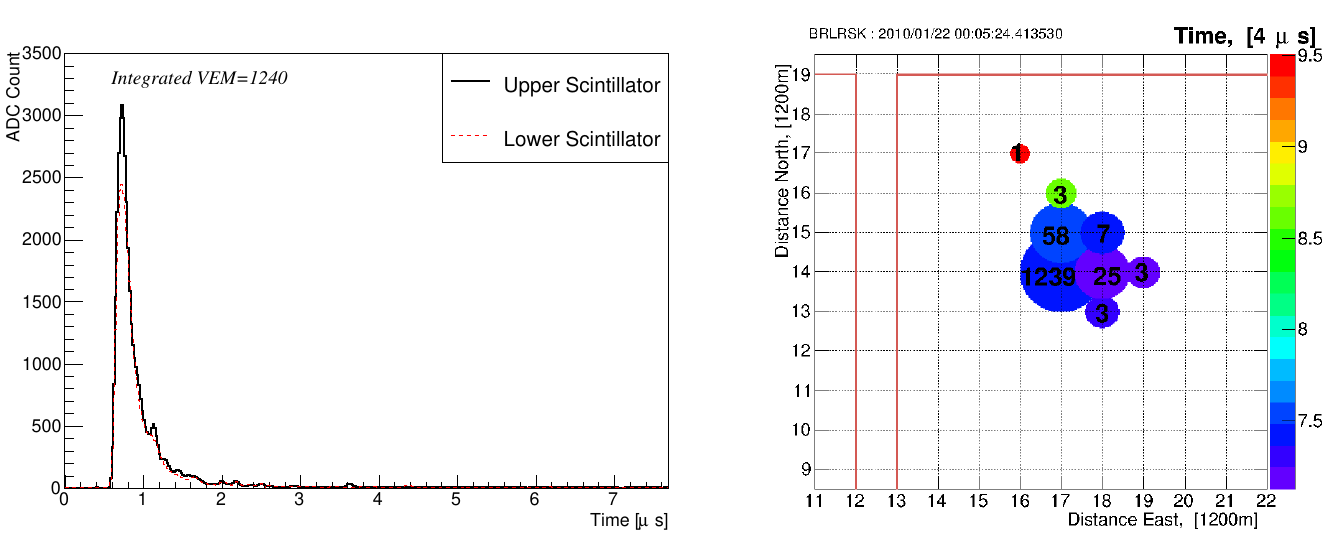}
\caption{{\em Left:} Upper and lower scintillator waveforms in a single surface detector unit, for a cosmic ray event, with an energy of 12.2~EeV and a zenith angle of $23^{\circ}$, observed at 00:05:24 on 22 January\ 2010.  {\em Right:} Footprint of TASD hits for all detectors units involved in the cosmic ray event, with the numbers indicating the Vertical Equivalent Muon (VEM) counts (see text), and the color indicating the relative arrival times.} 
\label{fig:crexample} 
\end{figure}

The surface detector photomultiplier waveforms recorded in the lightning-correlated bursts are different from those observed for cosmic rays, shown in Figure~\ref{fig:crexample}. In a typical cosmic ray event, the waveforms for SDs near the core (center-of-energy deposit) of the event are characterized by a single sharply defined shower front followed by a tail consistent with the expected shower thickness for a log normal-like signal. The upper and lower scintillator waveforms are well-matched~\citep{AbuZayyad:2012ru}. In contrast, the waveforms recorded during lightning bursts (as shown in Figure~\ref{fig:2figsB}) tend to show a slower rise and fall in the signal waveform. Also, the upper and lower scintillator waveforms are less-well matched, indicating that many of the detected particles -- primarily Compton electrons produced within the detector structure -- are detected only in one or the other of the two scintillator planes.  The extent to which the upper and lower signals are correlated is an indicator of the electrons having sufficient energy to propagate through both scintillators (see Discussion).

\begin{figure}
\includegraphics[width=\textwidth]{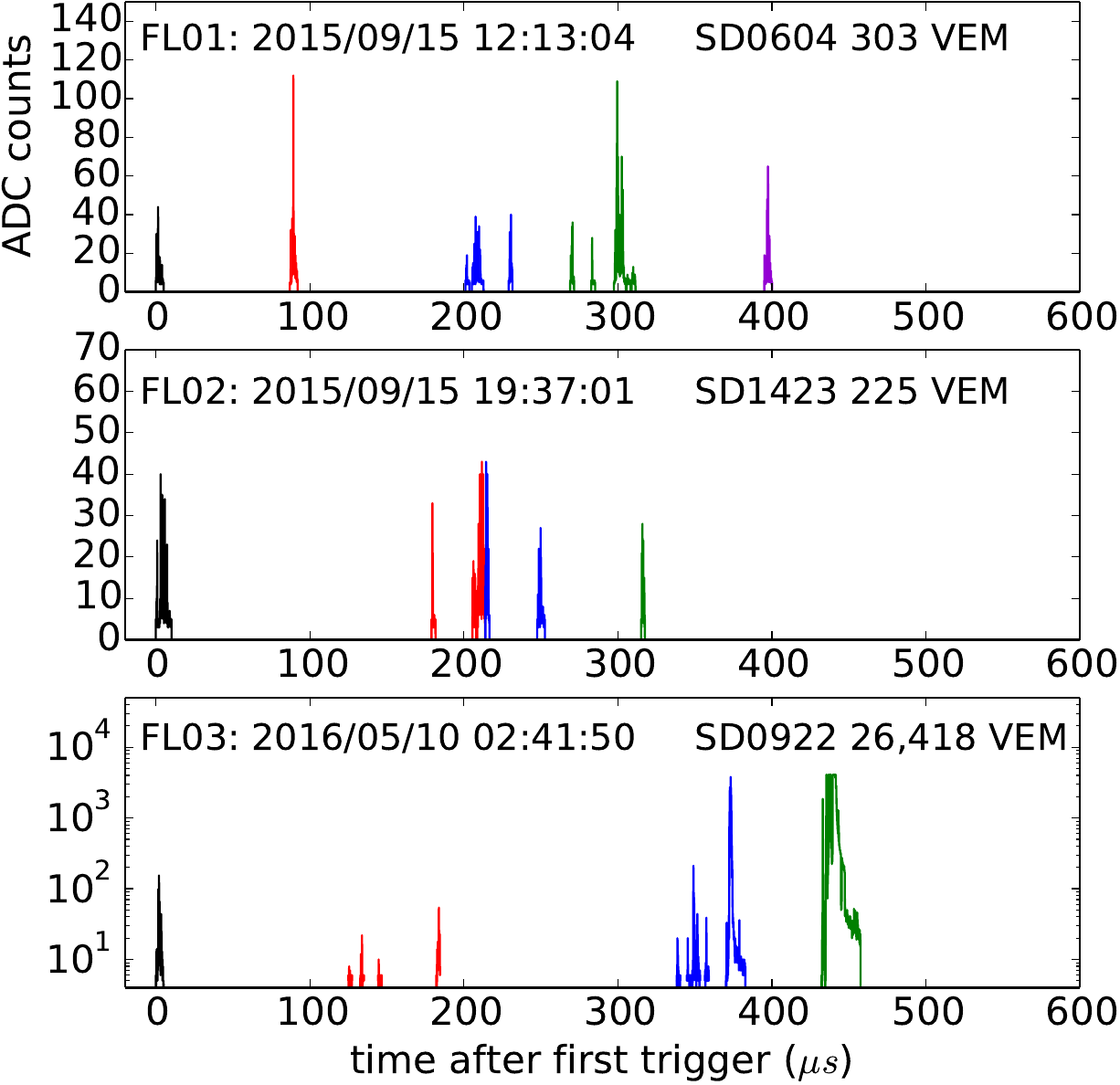}
\caption{{\em Top}: the combined waveform, for one of the SD (XXYY=0604), for all five triggers observed in the burst FL01. Each trigger in the burst is colored individually. Triggers 1 through 5 are colored in black, red, blue, green and violet. {\em Middle}: same, except at SD (XXYY=1423) for the burst of four triggers observed in FL02. {\em Bottom}: same, except at SD (XXYY=0922), for the burst of four triggers observed in FL03. The waveforms are found to be temporally resolved into discrete components, most of which are less than 10 microseconds in duration and which occur in succession over a duration of a few hundred microseconds. 

(The corresponding figures for the slow antenna-correlated showers are included as Figures S11 and S12 in the Supporting Information.)} 
\label{fig:WFS} 
\end{figure}

\clearpage

\subsection{LMA-Correlated Photon Showers}
\label{sec:lmacorr}

\begin{figure}[t]
\begin{center}
\includegraphics[width=\textwidth]{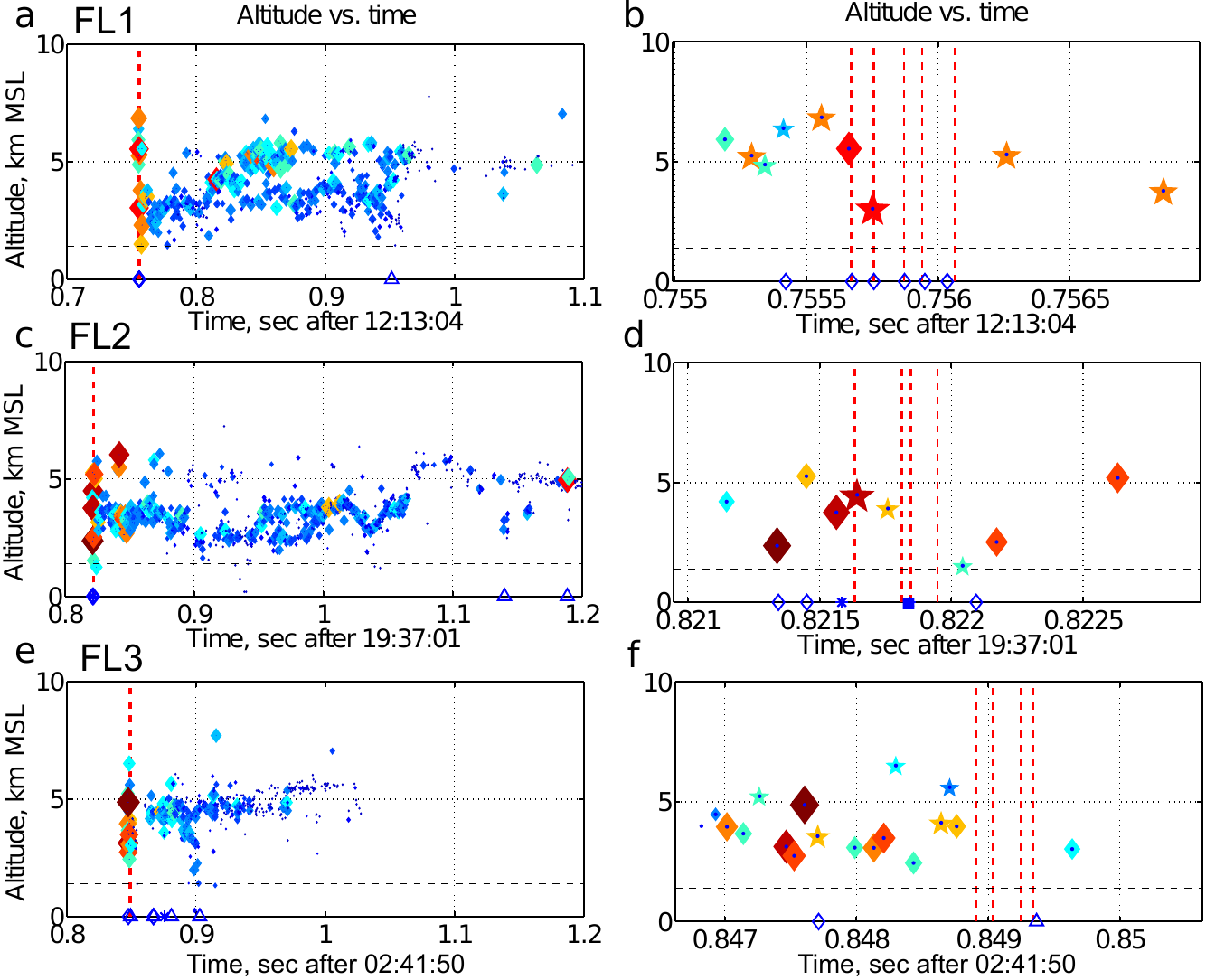}
\caption{Observations of the LMA-correlated  energetic radiation burst , showing altitude versus time plots of the LMA sources (colored diamonds) and the TASD trigger times (red dashed lines).  The left panels show the complete flashes and the right panels show zoomed-in views during the first 2--3~ms of each flash.  The LMA sources are colored and sized by the log of their radiated power, and range from source powers of --20~dBW (10~mW; blue colors) up to $+$25~dBW (320~W; red colors). NLDN events are shown on the abscissa: $\diamond$ = --IC, {\scriptsize $\triangle$} = --CG, {\scriptsize $\blacksquare$} = +IC, {\large $\ast$} = +CG. The mean altitude of the Telescope Array was $\sim$1.4~km above Mean Sea Level (MSL) (horizontal dashed line).  In the zoomed-in plots, a number of the altitude values are in error due to including sources having relaxed $\chi^2$ goodness of fit values, or being RF-noisy higher power events. Sources in the left panels have $\chi^2_\nu$~$\le$~5; starred sources in the right panels have $\chi^2_{\nu}$ values between 5 and 500. In Panel d, we deduce from the neighboring LMA sources that the the NLDN event identified as $+$CG is actually a misidentified $+$IC (See also Table S1a,b). }
\label{fig:new_fig3}
\end{center} 
\end{figure}

The LMA coverage was intermittent during the two-year period between 2014 and 2016. However, three energetic radiation bursts were observed in coincidence with LMA activity during this time. Figure~\ref{fig:WFS} shows composite TASD waveforms for each of the LMA-observed flashes.  The waveforms are found to be temporally resolved into discrete components, most of which are less than 10 microseconds in duration. These components occur in succession for a duration of a few hundred microseconds.

Figure~\ref{fig:new_fig3} shows how the TASD trigger times were related to the LMA source heights versus time, and to the NLDN observations. In each case the triggers (dashed red lines) occurred within the first 1-2 ms of the flash, relative to the time of the first LMA source. Except for the requirement that the showers be detected at three adjacent stations at 1.2~km spacing, the lack of TASD activity after the first 1--2~ms of the discharges cannot  be an artifact of the TASD trigger and data acquisition system, which continually processes high rate activity at the microsecond scale in order to efficiently identify cosmic ray air showers. 

Flashes FL1 and FL2 occurred in different storms on 15 Sept 2015. In both cases, the LMA observations (Figures S6 and S7) show that the flashes began as low-altitude intracloud discharges between mid-level negative charge at 5-6 km altitude MSL (3.5-4.5 km AGL), and deep, widespread lower positive charge at 2-4 km altitude MSL (0.5-2.5 km AGL). These are unusually low altitudes, which are typically $\sim$2~km higher in normally electrified convective storms. In addition, the flash rates were low, respectively producing only 2 and 5 flashes in 10 minute intervals around FL1 and FL2. The storms were thus not convectively vigorous, but the quiescent intervals leading up to FL1 and FL2 were noticeably long (4-5 minutes), likely allowing the electric forces to build up to strong values before the flashes were triggered. Once triggered, the flashes spent 200 and 300 ms discharging the large regions of lower positive charge before producing negative strokes to ground.

The LMA sources in the Figure~\ref{fig:new_fig3} plots are colored and sized by their VHF source powers.  The left-hand panels show that the highest-power sources (red colors; 15 to 25~dBW; 32--320~W) occurred at the beginning of the flashes, in conjunction with the TASD triggers.  Source powers for the remainder of each flash ranged from $\simeq$15~dBW down to minimum detectable values of $-$20~dBW (10~mW; blue colors).  The zoomed-in plots of the right-hand panels show the first 2-3 ms of the flashes in more detail. A number of the TASD triggers (red-dashed lines) were correlated with LMA and/or NLDN events.  The observations are tabulated to the microsecond level in Table S1 of the supporting information, where correlated events are highlighted. It should be noted that the altitude of some of the LMA sources are in error (by a couple of km) due to the VHF radiation being non-impulsive and continuously noisy around the time of its peak. Mis-locations are a characteristic feature of some higher power sources that is useful in identifying their noisy nature, as found in the study of NBEs and fast positive breakdown by~\citep{rison2016}. Despite being mis-located, the timing and source powers of the events are reasonably well-determined, however, as indicated by the events being correlated in time with the TASD triggers.

From the zoomed-in LMA observations,  the heights of the TASD trigger events for FL1 and FL2 can be estimated quantitatively by averaging the altitudes of the the LMA events leading up to the triggers. For FL1, the initial six LMA sources had an average height $z = 5.8 \pm 0.7$~km MSL (4.4~km AGL). For FL2, ignoring the three obviously incorrect low-altitude sources in the zoomed-in panel, the average initial activity was at a lower altitude than for FL1, $z = 4.4 \pm 0.6$~km (3.0 km AGL).

The TASD triggers occurred within the first ms of each flash as the initial negative breakdown descended toward and into the lower positive charge region. Also, the TASD stations that were triggered by the radiation showers, as well as the corresponding NLDN sources, were located directly below the initial LMA sources (Figures~\ref{fig:2figsB} and S1), consistent with the radiation being produced by downward-directed negative breakdown.

Flash FL3 occurred on 10~May~2016. It was unusual in that its initial
leader went directly to ground at a high speed, reaching ground 2.6~ms
after flash initiation (as determined by the first LMA source time;
Table~S1b and Figure~S8). Assuming the leader was initiated at 5~km
MSL altitude, its average 1-D propagation speed to ground was $1.4
\times 10^6$~m/s, an order of magnitude faster than the speed of
typical stepped leaders. The ensuing return stroke was similarly
energetic, having a peak current of $-$94~kA.  The first TASD trigger
occurred 2.1~ms after the flash initiation, relatively late in
comparison to the total leader duration. The final trigger occurred
just 26~$\mu$s before the start of the return stroke. From this, we
can deduce that the first TASD trigger occurred when the leader was
$\simeq$640~m above ground, and the final trigger occurred when the
leader was $\simeq$40~m above ground. The VEM count in the nearest
surface detector was at least 22,000 (45~GeV energy deposition) for
the last TASD trigger. This compares to VEM values of a few tens to a
few hundreds for other events of this study (Table S1), and is likely
due to the lower source altitude. Despite their lower altitude, the
FL3 triggers had similarly-sized footprints on the ground (Figure S1)
as the higher altitude events of FL1 and FL2. 

%In Section~\ref{sec:discussion}, we present the results of a GEANT4~\citep{Agostinelli:2002hh,Allison:2006ve} simulation explaining these observed ground-level gamma fluxes in terms of the flux at the source.

\subsection{Slow Antenna-Correlated Photon Showers}

Slow electric field change data were collected only during the 2014 storm season. Seven TASD trigger bursts were correlated with the observations (Table S2, S3, and Figures S2-S4, and S11-S12). The flashes occurred on three different days in 2014, during times when LMA data was not available. Nevertheless, the slow antenna waveforms are readily understood~\citep{krehbiel1979}.  As seen for flashes FL1 and FL2 of the LMA correlations, the TASD trigger bursts occurred in the first millisecond of the flash, relative to the start of the slow antenna field change, for each of the seven events.  

Two events (flashes FL4 and FL8), had normal--duration initial leaders
(28~ms and 14~ms) and typical return stroke peak currents (--12.2 and --11.7~kA). Of the remaining five events, three had intermediate--duration initial leaders between 4 and 10~ms (FL7, FL9 and FL10; $I_{\rm pk}$ = --66.6, --16.1, --35.0~kA, respectively). The remaining two (FL5 and FL6) had fast, short--duration leaders (1.5--2.5~ms) and correspondingly strong return strokes (--140~kA and --101~kA peak). Thus the TASD triggers were not exclusively associated with energetic leaders.

\begin{figure}
%\vspace{-2\baselineskip}
\includegraphics[width=\textwidth]{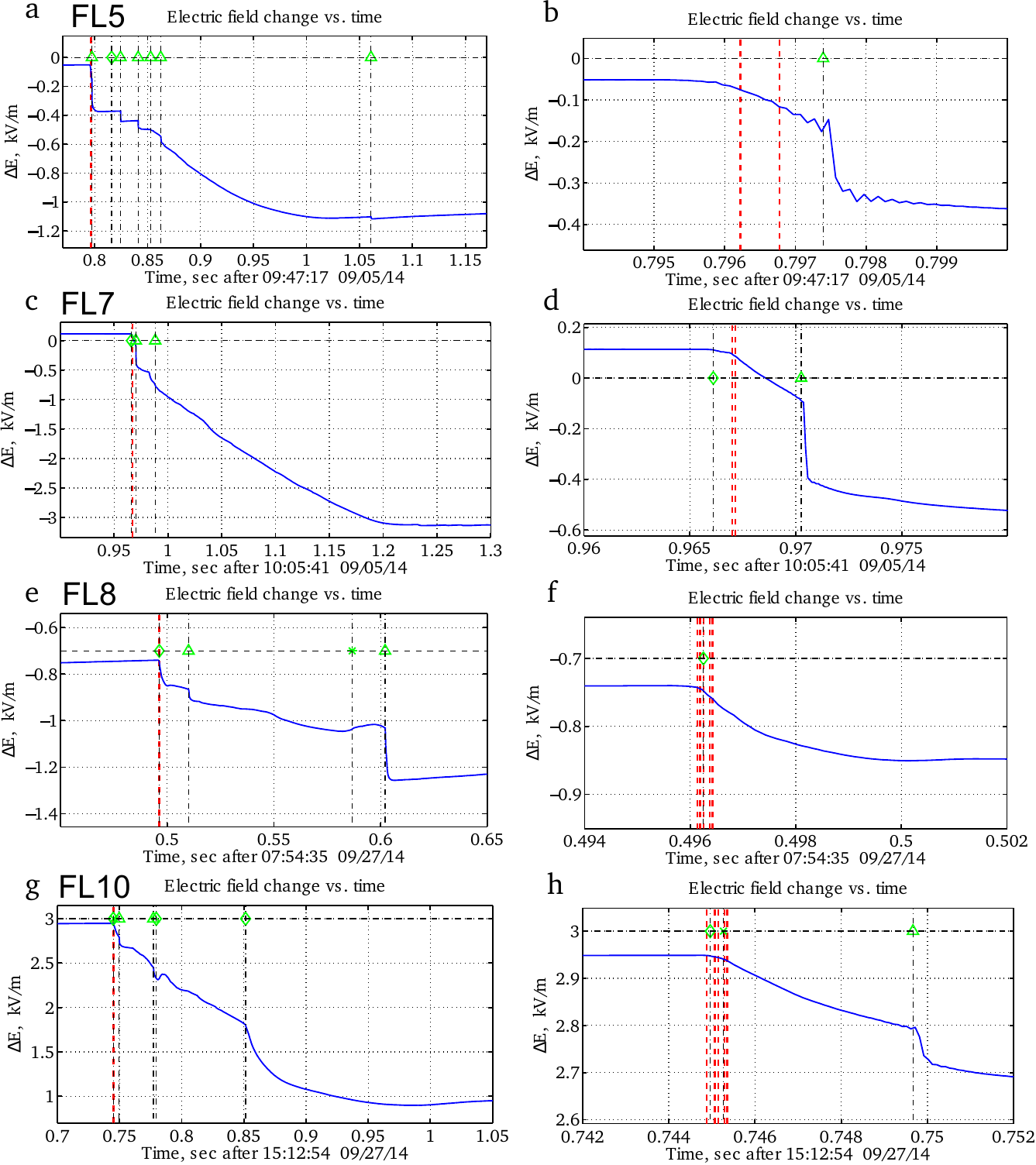}
\caption{Electric field change versus time for four of the slow-antenna correlated trigger burst events (Flashes 5, 7, 8 and 10).  The dashed red lines show the TASD trigger times, while the dot-dashed lines and green symbols indicate the times and type of NLDN events ({\scriptsize $\triangle$}, $\diamond$, $\times$,{\large$\ast$} = --CG, --IC, $+$CG, $+$IC, respectively). The long negative field changes at the end of flashes are continuing current discharges to ground. The left side panels show the entire flash, while the right side panels zoom in on the initial 5 to 20 ms of the flash.} 
\label{fig:slow_a}
\end{figure}

The observations are presented in Figure~\ref{fig:slow_a}, which shows
correlation results for flashes 5, 7, 8 and 10.  (The remaining
flashes are shown in Figure S5). The overall $\Delta E$ waveforms are
characteristic of multiple-stroke negative cloud-to-ground discharges
(left panels). Of interest here is the duration of the radiation bursts
relative to the leader duration (right panels). As can be seen
qualitatively from the plots, except for FL5, the burst durations are
a relatively small fraction of the leader duration. The fractional
durations for FL7, FL8, and FL10 of Figure~\ref{fig:slow_a} are
$\sim$3.4\%, 2.1\%, and 10\%. Assuming for simplicity that the leaders
were nominally 5~km in length and propagated to ground at a constant
speed, the above percentages indicate spatial extents of 170~m, 100~m
and 500~m respectively during the burst intervals. For comparison, the
stepping length of upward negative breakdown at the beginning of
intracloud flashes is $\sim 500$~m (e.g. \cite{rison2016}). Because the
stepping lengths will be shorter at the lower altitude (higher
pressure), the above values are qualitatively consistent with the
bursts being associated with one or two individual steps of the initial breakdown of the flash. The fractional duration for FL5 is approximately 22\%, indicating that the TASD triggers are probably indicative of  separate events during the leader descent.

\subsection{Optical and Radio Frequency Interference}

The TASD is designed for 24-hour operation in a high-desert environment, between typical low temperatures of $10^{\circ}$~F and high temperatures of $100^{\circ}$~F. In eight years of operation in this environment, the system has proven robust against temperature and light-level fluctuations (including in broad daylight) as measured by once-per-minute pedestal and count rate samples on each of the 507 surface detectors~\citep{AbuZayyad:2012kk}.

The detector scintillator, photomultipliers and data acquisition electronics are enclosed in electrically sealed and grounded steel boxes in order to shield against RF interference. The system has been proven to be generally adequate for cosmic ray detection, however it is reasonable to be concerned with system performance in the extreme thunderstorm environment being considered in this paper. Thus the waveforms of all SDs included in this study were examined for evidence of electrical or optical interference. 

The only observed effect of an electrical transient is seen during the final trigger of FL3, at the station that experienced the 22,000~VEM radiation event (surface detector SD0922, bottom panel of Figure~5 and Table S1a).  The trigger occurred 26~$\mu$s before the leader connected to ground, corresponding to an estimated $\simeq$40~m above ground. From the NLDN data, the ensuing return stroke had a peak current of $-$94~kA and was located $\simeq$80~m north of SD0922. The footprints and detailed waveforms for each of the four triggers at SD0922 are shown in Figures~S13 and S14, and waveforms at each of the four stations that detected the 4th trigger are shown in Figure~S15. The lower right panel in Figure~S14 and the lower left panel in Figure~S15 shows that the 4th trigger strongly saturated the SD0922 waveform.  While much of the high particle flux is due to the proximity of the gamma source (likely less than 100 meters) to the detector unit, the waveform shows two abrupt amplitude decreases as the signal died out.  The 2nd decrease, at $\simeq$15~$\mu$s in the record, corresponded to the time of the NLDN-detected return stroke and persisted, while the first decrease 2~$\mu$s earlier abruptly recovered after $\simeq$1~$\mu$s.  In both instances, the changes were due to abrupt decreases in the signal pedestal (electrical ground) (see logarithmic plot of Figure~S14). 

The other surface detectors involved in the fourth trigger show no anomalous behavior (Figure~S15), nor does SD0922 for the first three triggers of this burst (Figure S14). Thus while RF transients can potentially produce waveform interference in the case of direct or nearly-direct hits of SDs by lightning, we determine that this is not a pervasive problem in this study and that it has no impact on the conclusions drawn. 

Similarly, there is no indication of optical interference at the TASD stations, even during the exceedingly bright return stroke that would have occurred close to SD0922, or of any triggers being produced by other flashes that had strong return stroke currents.

\clearpage

%%%%%%%%%%%%%%%%%%%%%%%%%%%%%%%%%%%%%%%%%%%%%%%
%\input{discussion_grl.tex}
\section{Discussion}
\label{sec:discussion} 

\subsection{Characteristics of Observed Events}

In this paper we report observations of energetic radiation bursts produced during downward negative breakdown at the beginning of low-altitude cloud-to-ground and intracloud flashes. The bursts occurred during the first 1--2 ms of the discharges and have overall durations between 87 and 551~$\mu$sec. With the high-resolution timing of the TASD, the bursts are found to consist of several (2--5) individual components, each of which are a few microseconds in duration, separated in time by $\simeq$10-250~$\mu$s between events.

It is typical for negative CG discharges to begin with downward negative breakdown, particularly in the initial millisecond or more of the discharges. This is shown by the VHF sources for the LMA-correlated events, and is corroborated by the TASD footprints, which show that the cores of the energetic showers were directly below the initial LMA sources (Figures~\ref{fig:2figsB} and S1). Similarly, the NLDN cloud events at the beginning of the flashes (Figure~\ref{fig:new_fig3} and Table~S1) agree with the LMA and TA observations. 

Although the electric field change (SA) observations do not locate the flashes, the NLDN data shows their locations, and similarly agree with the  locations and times of the TASD events.  In addition, the SA leader field changes are the same polarity as the ensuing return strokes, namely negative, consistent with downward negative leader breakdown for flashes beyond the reversal distance. It is also worth noting that even though there is no direct altitude data for the SA-correlated flashes, both the LMA and SA data show that the TGF events occur in the first 1-2 ms of the flashes, and the TASD footprints are comparable for the LMA and SA events. It is therefore reasonable to assume that the SA-observed TGF events have similar source altitudes as for the flashes observed with the LMA.

The reported observations were facilitated by the fact that the events occurred over a relatively high-altitude land, averaging approximately 1.4 km MSL. This, coupled with the events being downward-directed, and originating at low altitude (2.5-4.5 km AGL), minimized the atmospheric attenuation losses.

The principal question is whether the observed showers are indeed comprised substantially of gamma-rays --- usually defined as having energies in the nuclear decay range of order 100~keV or greater --- or whether the predominant source of energy deposit in the TASD is due to lower-energy $x$-rays. 

A simple argument can be made on the basis of the waveforms, for example Figure~\ref{fig:2figsB} (or supporting figures S1--S4). In each case, the waveforms consist of contributions from individual Compton electrons. For cosmic ray events, a single vertical equivalent muon, or 2~MeV energy deposit, produces a change of $\simeq 30$~counts in the ADC signal, lasting about 100~ns. One ADC count therefore corresponds to (1/30) of a VEM, or $\simeq 70$~keV. Electrons with this kinetic energy require a photon of at least 170~keV to be produced by Compton scattering (Figure~S9), corresponding to the minimum detectable ADC signal. From Fig.~3, the individual pulses at the beginning of the waveform (between 7.5 and 10~$\mu$s) produce signals ranging from about 5 up to 30~ADC counts above background, corresponding to electron energies of 350~keV and $\simeq$2.0~MeV, respectively, and minimum individual photon energies of 0.52~MeV up to 2.2~MeV (Fig.~S9). (It is important to note that minimum ionizing Compton electrons above 2~MeV will deposit no more than 2~MeV per 1 cm scintillator plane, although they may deposit that amount of energy in both planes. In terms of the photon energies required to produce Compton electrons of 2~MeV (or stronger), the 2.2~MeV minimum value corresponds to the photon being backscattered. At grazing incidence, higher energy photons are needed, with the most probable photon energy for a given electron kinetic energy being larger by factor of $\simeq$3, or 6.6~MeV (Figure~S10).) The subsequent, larger amplitude signal in Figure~\ref{fig:2figsB} consists of multiple photons of similar or potentially larger energies, with the overall event having a total VEM count of 227~VEM (450~MeV). This is well above the trigger threshold of 3~VEM within 8~$\mu$s, and the photon energies are well within the range of gamma rays.

Extending the simple argument further, we refer to the GEANT4 simulations of Figure~2. The upper panel shows the average energy deposited in each of the scintillators by Compton electrons generated in the air just above the surface of a TASD station. The energy deposition shows a sharp increase for electrons above 10 MeV, and no energy contribution below a few MeV. The latter result is due to the energy loss in the weather shield and uppermost shielding box being about 1.9 and 2.4~MeV, respectively, so that Compton electrons produced in the air above the detector with energies less than about 4.5~MeV will not penetrate down to the top scintillator.  Lesser energy electrons would have to be produced within or below the upper level steel or in the scintillators to contribute to the detector signal.  On the other hand, 10~MeV electrons would make it all the way through the steel and both scintillators, depositing about 2~MeV in each. That the latter occurs is demonstrated by the fact that the VEM counts in the upper and lower scintillators are not completely random, but are partially correlated.  For a Compton electron to penetrate completely through both scintillators, its energy would need to be greater than $2 \times 2$~MeV + 1.6~MeV loss in the 1~mm stainless separator, or $\geq 5.6$~MeV upon entering the top scintillator.  This corresponds to a minimum Compton photon of 5.85~MeV. Compton electrons less than 5.6~MeV would be partially transmitted through one or both scintillators depending on where they are generated, and may partially contribute to the top/bottom correlation.  For Compton electrons generated in the air above the weather shield, their energy would have to be greater than $5.6 + 4.5 \simeq 10.1$~MeV to penetrate down through both scintillators, consistent with the simulation results in the lower panel of Figure~\ref{fig:depvalt}. Taken together, the extent to which the upper and lower scintillator counts are correlated is further evidence that the shower photons are in the multi-MeV range of energies.

Other considerations include the effects of attenuation and scattering in the atmosphere above the TASD. The absorption length of photons in the atmosphere at TA altitudes is of order 100~m at 100~keV, and less than 10~m at 10~keV~\citep{Olive:2016xmw}. Because the leaders associated with the TASD occur at up to 4~km above ground level (Section~\ref{sec:lmacorr}), we can expect that substantial attenuation of lower-energy photons will occur in the atmospheric overburden. 

In order to be more quantitative, we performed a GEANT4 simulation incorporating a model of the atmosphere as well as the TASD (Section~\ref{sec:ta}). We varied the energy and altitude of the primary photons, and recorded the total energy deposited in two scintillator planes placed in front of and perpendicular to the photon beam. The layers of steel in the TASD were also included. For this particular simulation, the ``detector'' was made sufficiently large that losses due to the horizontal development of the showers were (conservatively) ignored. 

\begin{figure}
\begin{center}
\includegraphics[width=\textwidth]{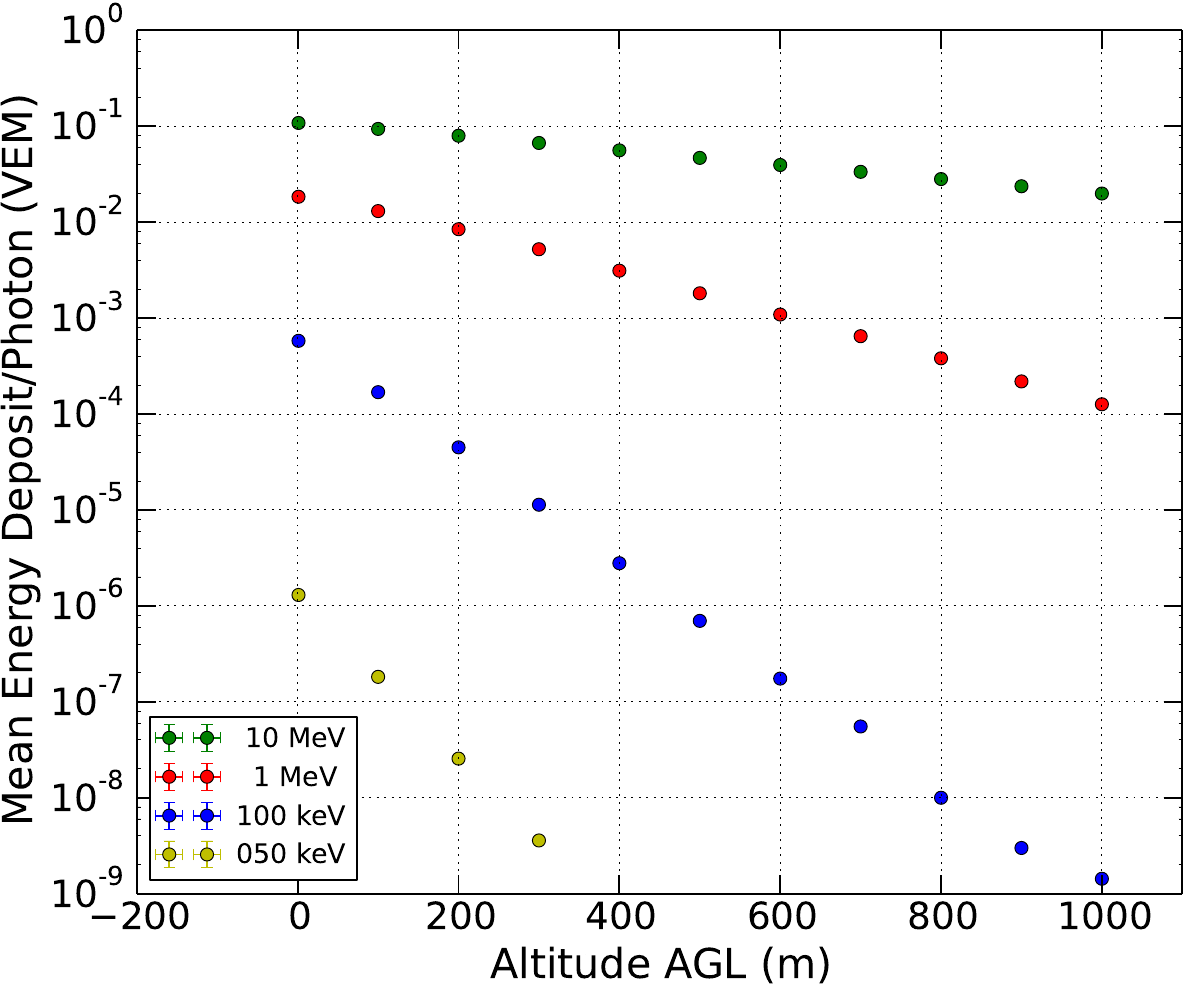}
\caption{Mean energy deposit (average of two TASD planes) per primary photon, versus altitude AGL for various photon energies. For this simulation, losses due to shower lateral spread are ignored.} 
\label{fig:edep}
\end{center}
\end{figure}

The results of this simulation are summarized in Figure~\ref{fig:edep}, showing the mean TASD energy deposit versus altitude for various primary photon energies. The effect of decreasing photon attenuation in the $x$-ray to gamma-ray transition range is significant: At an altitude of 1~km AGL, the mean energy deposited by a 1~MeV photon is a factor of $10^{5}$ greater than that of a 100~keV photon. For reasonable energy spectra (see {\em e.g.} Figure~\ref{fig:new_fig3} of \cite{Dwyer2012}) the corresponding decrease in flux is far less, only one or two orders of magnitude. Thus from the GEANT4 simulation for sources at an altitude of 1~km or more -- FL1 and FL2, and likely most of the slow-antenna events -- the TASD signal is due to primary photons with energy of order 1~MeV or greater.

Lower altitude sources, such as those seen in FL3 with a
short-duration energetic leader, may have  significant contributions from lower energy primary photons
if the source is sufficiently close to the detector unit. However, the
last and most energetic FL3 burst triggered four TASD units, implying
that the shower must  also have had a significant
gamma-ray component  in addition to x-rays, both for the final and earlier triggers
  of FL3. This differs with interpretations of the similar low altitude event
  observed by~\cite{moore2001} as being caused only by lower-energy x-rays
  (e.g.,~\cite{dwyer2004a}).

  Although the 22,000 VEM count (45 GeV) of the low-altitude trigger of FL3
  may seem like a large amount of energy, it is minuscule in comparison to the
  total electrostatic energy available to the leader that produced the
  trigger.  In particular, the amount of charge $\Delta Q$ lowered to ground
  by negative stepped leaders similar to that of FL3 is typically
  $\simeq$5-10  Coulombs \citep{Krehbiel:1981thesis}.  This occurs across a potential
  difference $\Delta V = 100-200$ MV between the negative charge region of the
  cloud and ground (e.g.,~\cite{RU2003, krehbiel2008}). The
  amount of energy available to the leader discharge processes is given by $W
  = \Delta Q \cdot \Delta V \simeq 10^9$~Joules.  The energy deposited in the
  TASD detector by the final burst of Flash 3 was about 45~GeV, or $45 \times
  10^9$~eV.  Since 1~eV~$= 1.6 \times 10^{-19}$~J, 45~GeV corresponds to $7.2
  \times 10^{-9}$~J, or $\simeq 10^{-8}$~J. This will be some fraction
  of the
  total energy of the radiation burst, but is still only a minuscule fraction
  of the overall energy available for the leader.

\subsection{Fluence Estimates}
With the gamma-ray component of the showers established,  in this section we
investigate whether the radiation bursts could be produced by photon showers with energy spectra similar to that observed in upward-pointed TGFs. In a variant of the simulation study described above, photons were generated from a pointlike source, according to a RREA spectrum 
\begin{equation}
\label{eq:rrea}
\frac{dN}{dE} \sim \frac{e^{-E/(7~{\rm MeV})}}{E}
\end{equation}
where $E$ is the photon energy above 10
keV~\citep{Xu:2015thesis}. The primary purpose of
the simulations is to estimate the source fluences for comparison with similar
estimates obtained from satellite observations, which are obtained in the
same manner assuming a RREA spectrum.

Based on our observation of the shower footprint size and leader altitudes, particles are assumed to be forward beamed within a cone of half-angle 16$^{\circ}$. The angular distribution of particles is assumed to be isotropic within that cone. Particles are then tracked through the atmosphere and TASD detector model. 

\begin{figure}
\begin{center}
\includegraphics[width=\textwidth]{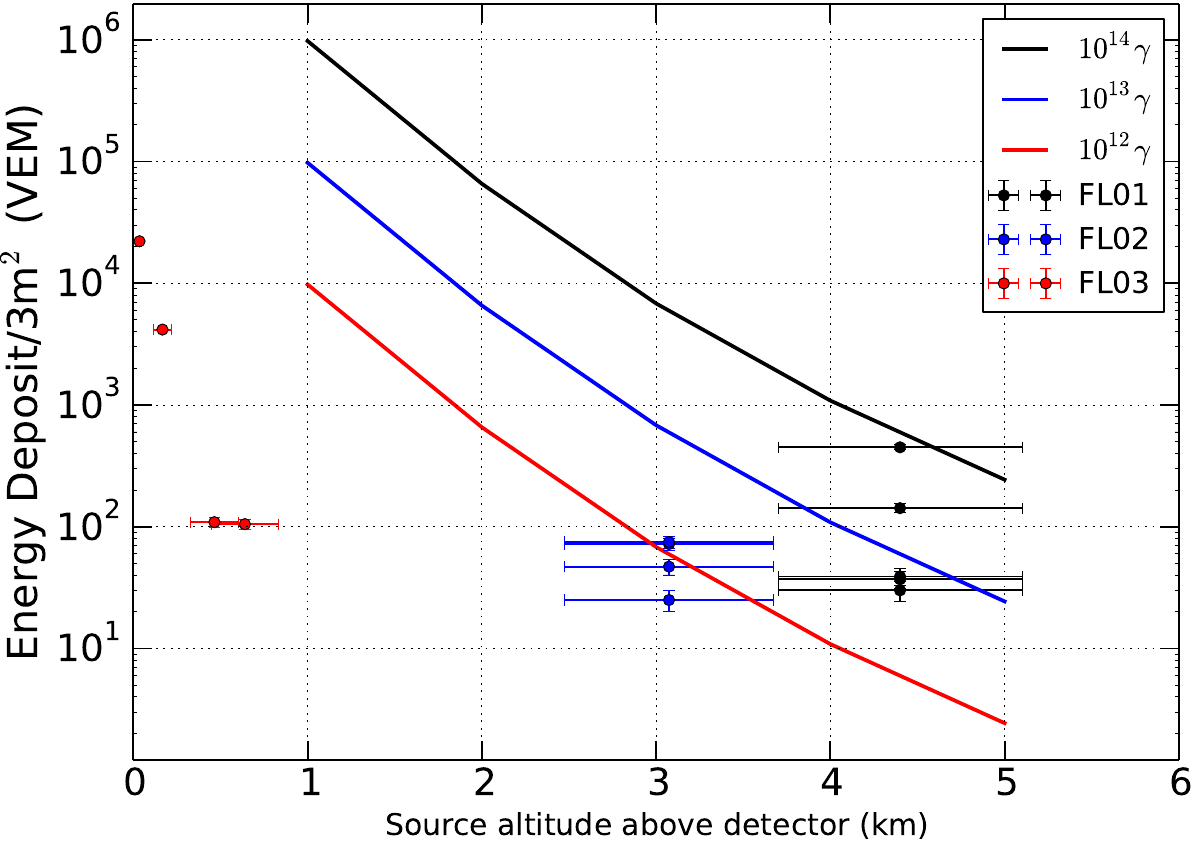}
\caption{ Energy deposit (Vertical Equivalent Muon units) in a TASD scintillator in the path of a downward-directed RREA shower, as a function of shower initiation altitude. Red, blue and black curves represent $10^{12}$, $10^{13}$ and $10^{14}$~gamma ray primaries, respectively. Details of the RREA simulation are given in the text. Data for events FL01-FL03 are superimposed, with estimated uncertainties.}
\label{fig:depvalt2}
\end{center}
\end{figure}
Figure~\ref{fig:depvalt2} summarizes the results of the simulation, showing the energy deposit in VEM units in a 3~m$^2$ TASD as a function of the altitude of the point source above the detector.  It does this for source fluences of $10^{12}$, $10^{13}$, and $10^{14}$ primary photons. For an event at 3~km altitude AGL, corresponding to flash FL2, the observations are consistent with a source of $\simeq$10$^{12}$ RREA photons. For flash FL1, initiated an estimated 4.4~km AGL, the observations are consistent with a maximum fluence of $\simeq$10$^{14}$ photons and a minimum fluence of less than $\simeq$10$^{13}$ photons.

The isolated red dots in Figure~\ref{fig:depvalt2} show the
observations for the radiation events of flash FL3. Due to the leader
reaching ground in 2.6~ms, and the radiation events occurring
relatively late in the leader, the events originated at much lower
altitudes, starting at $\simeq$640~m and $\simeq$470~m AGL for
  the first two triggers, with energy
depositions of $\simeq$100~VEM, and
finishing with rapidly stronger emissions at 170~m and 40~m AGL
  altitude, with the latter producing a highly energetic VEM of
$\simeq$22,000. The latter
  two triggers could not have been produced by a large-scale RREA process, as
  the events originated too close to the ground for the avalanching to fully
  develop.  Nevertheless, to the extent that a RREA spectrum
applies to the lower altitude emissions, the final two events would
have fluences of $\simeq 10^{10}$ -- $10^{11}$ photons, while the two
initial events would have been one or two orders of magnitude weaker
yet. In any case, the observation that substantial gamma
  radiation can be produced as the leader descends toward ground,
  where the stepping process is over noticeably shorter
  distances ($\simeq 50$~m, e.g.,~\cite{RU2003}) than in 
  the initial breakdown stages, indicates that high energy gamma rays
  can be produced by relatively short-scale avalanche processes.
%Although the same arguments apply for these events having a substantial fraction of high energy gamma radiation (particularly for the final, low-altitude event, given that it was detected at 3 outlying stations), the fact that the total production is less suggests a modified process. This is not surprising, as the stepping process later in stepped leaders to ground is noticeably different (shorter distances and times between steps) from that at the beginning of the downward breakdown.

\subsection{Comparison with Other Observations}

The downward TGFs of this study are similar to satellite-detected TGFs in that satellite events which can be correlated with ground-based sferic observations are typically found to occur in the beginning stages of negative polarity breakdown (e.g. \citep{Stanley2006, cummer2011, cummer2015}). In both cases the overall durations are also similar, lasting up to 500~microseconds or longer. However, the observations differ in that a) the downward TGFs of the present study consist of a sequence of a few isolated and relatively short-duration bursts, whereas the satellite-detected events are more continuous with time over the full duration of an event. More significantly, b) the fluences of the satellite TGFs are substantially larger than those of the present study, with estimated values of $\simeq 10^{16}$ to $10^{18}$ primary photons~\citep{smith2011b, JGRA:JGRA21798}, two to four orders of magnitude larger than the maximum estimated fluence of the present study.

As discussed below, the burst-like nature of the gamma radiation seen
in the present study is consistent with being produced by the stepping
process of negative-polarity breakdown.  In addition, the observations
are in full agreement with the assessment by~\cite{celestin2012} 
  and ~\cite{daSilva2015}  that the longer duration $\ge$100~$\mu$s TGF pulses of satellite observations could readily be due to overlapping emissions from much shorter ($\sim$10~$\mu$s) temporal durations.  The increased duration is caused by the effects of Compton scattering, which lengthens the possible paths between the source and detector.  The effects of scattering and of resulting overlapping emissions can be seen for example in Figure~2 of the study of Fermi data by~\cite{foley2014}.

Due to the much shorter path lengths between the source and detector for ground-based measurements ($\simeq3$--4~km, vs. 600~km or more for satellite observations), ground observations provide a much clearer picture of the temporal production of gamma rays inside storms.

Concerning the 2--4 orders of magnitude difference in the fluence values, there are several explanations for the difference.  The first is that the simulations of Figure~\ref{fig:depvalt2} assume scattering in a cone having a half angle of $16^\circ$.  If the actual half angle is $45^\circ$, considered to be a realistic value in evaluating in other simulations and for evaluating the satellite observations (e.g. \cite{GRL:GRL20344}), the number of source photons required to produce the same energy deposition in the TASD is increased by close to an order of order of magnitude, due to radiation being distributed over a larger solid angle.  Another contributing factor could be the effect of upward intracloud sources being at higher altitude in the atmosphere (10-12~km or higher), and therefore at lower pressure, allowing electron avalanches to develop over larger distances and become more energetic than at lower altitudes.

Finally, due to their large range (600--700~km) and small detection cross-sections, satellite observations are necessarily biased to the largest events.  For example, ongoing studies by \citep{lyu2015,lyu2016} and \citep{cummer2017} are increasingly showing that TGFs detected by the Fermi Gamma-ray Burst Monitor (GBM) are associated with energetic in-cloud pulses (EIPs) having extraordinarily large peak currents of 150-300~kA or more.  By contrast, the NLDN currents of IC events that were correlated with the TASD triggers in the present study ranged in magnitude from $\simeq 15$~kA to 35~kA peak (Tables S1--S3).  As a separate example of the insensitivity of satellite measurements, the GBM detects an average of $\sim$100 counts per TGF event (e.g. \citep{foley2014,mailyan2016}). If the average fluence of the events is $10^{17}$ photons, then TGFs that generate $10^{15}$ photons would be expected to result in detection of only a single photon, which would not be  distinguishable from noise by satellite observations.

Another type of gamma radiation is detected at the ground beneath
electrically active storms, called thunderstorm ground enhancements
(TGEs) (\cite{chilingarian2017} and references therein). These are a
class of weak gamma and x-ray ``glows'' that develop during
inter-flash intervals of storms (e.g.~\cite{kochkin2017}). The glows are produced by electron avalanches in localized regions of strong electrostatic fields in the storm, and are terminated by lightning flashes. Similar observations of x- and gamma-rays have been reported by \citep{eack1996,eack2000} from {\em in-situ} balloon-borne measurements inside storms.  Gamma-ray glows have also been detected by aircraft flying over the tops of electrically active storms~\citep{smith2011a,smith2011b,kelley2015,kochkin2017} In all cases the glows are a different phenomenon than TGFs and are even considered to be competing with lightning rather than being involved in its initiation~\citep{kelley2015}.

\clearpage

\section{Conclusion}
\label{sec:conclusion}

Taken together, the observations presented here provide a general description of downward-directed terrestrial gamma flashes associated with downward negative lightning leaders. The key points of these observations include:

\begin{itemize}

\item Bursts of gamma radiation observed on the ground occur during the first 1--2~ms of negative downward leader formation.

\item Burst durations are of order several hundred microseconds, and consist of several showers each lasting from roughly one to ten microseconds.

\item Showers whose sources are a few kilometers or less above ground level have footprints on the ground typically $\sim$ 3--5~km in diameter, and are capable of triggering at least three thin scintillator detectors on a 1.2~km grid within 8~$\mu$s. Thus the extent of the showers' propagation through the atmosphere indicates that photon energy must be in the gamma-ray regime.

\item Analysis of the scintillator waveforms show that Compton electrons produced by the showers have energies extending into the multi-MeV range, indicating the electrons are produced by photons in the gamma-ray regime.

%\item Simulation studies indicate that the photons in these showers must be in the MeV or greater energy range, at these altitudes, in order that they be detected on the ground. That, in addition to the footprint size, demonstrates that gamma radiation is a primary component of the observed showers.  

\item The observed energy deposit is consistent with forward-beamed showers of $10^{12}$-$10^{14}$ or more primary photons above 100~keV, distributed according to a RREA spectrum.

\end{itemize}

The result that the observations were confined to the first 1-2 ms of the discharges, and usually occurred in a single burst lasting a few hundred microseconds, suggests the TGFs were almost certainly produced by one or two particularly energetic leader steps at the beginning of the breakdown.  From this, the TGFs were produced by ``initial breakdown pulses'' (IBPs) at the beginning of IC and CG flashes (e.g., ~\citep{marshall2013} and Figs. 3 and S6 of~\citep{rison2016}). Such a correlation has been reported by ~\citep{lyu2016}, who compared three satellite detected TGFs with unusually high peak current (several hundred kA) NLDN events, which they termed energetic intracloud pulses (EIPs).  The present results are consistent with their finding, except that the NLDN currents are less strong (10-100 kA).  The actual correspondence of TGFs with fast electric field changes of IBPs remains to be demonstrated, however, and remains the subject of continued study.

Currently, both the LMA network and slow antenna electric field change instrument remain deployed at the Telescope Array site. An expansion by a factor of four in the coverage area of TASD is planned within the next several years. A plan is also in place to deploy additional slow as well as fast electric field sensors for improved coverage of the expanded TASD array. This will enable us to study the relation between SD observations and the development of negative breakdown in greater detail.  Combined with prolonged operation periods and continuous TA, LMA, and electric field observations, future studies will enable us to better identify and constrain the mechanisms of downward TGF production.

%%%%%%%%%%%%%%%%%%%%%%%%%%%

\section{Acknowledgements}

The lightning mapping array used in this study was developed and
operated with the support of NSF grants AGS-1205727 and
AGS-1613260. The Telescope Array experiment is supported by the Japan
Society for the Promotion of Science through Grants-in-Aids for
Scientific Research on Specially Promoted Research (15H05693) and for
Scientific Research (S) (15H05741), and the Inter-University Research
Program of the Institute for Cosmic Ray Research; by the U.S. National
Science Foundation awards PHY-0307098, PHY-0601915, PHY-0649681,
PHY-0703893, PHY-0758342, PHY-0848320, PHY-1069280, PHY-1069286,
PHY-1404495 and PHY-1404502; by the National Research Foundation of
Korea (2015R1A2A1A01006870, 2015R1A2A1A15055344, 2016R1A5A1013277,
2007-0093860, 2016R1A2B4014967, 2017K1A4A3015188); by the Russian
Academy of Sciences, RFBR grant 16-02-00962a (INR), IISN project No. 4.4502.13, and Belgian
Science Policy under IUAP VII/37 (ULB). The foundations of Dr. Ezekiel
R. and Edna Wattis Dumke, Willard L. Eccles, and George S. and Dolores
Dor\'e Eccles all helped with generous donations. The State of Utah
supported the project through its Economic Development Board, and the
University of Utah through the Office of the Vice President for
Research. The experimental site became available through the
cooperation of the Utah School and Institutional Trust Lands
Administration (SITLA), U.S. Bureau of Land Management (BLM), and the
U.S. Air Force. We appreciate the assistance of the State of Utah and
Fillmore offices of the BLM in crafting the Plan of Development for
the site.  We also wish to thank the people and the officials of
Millard County, Utah for their steadfast and warm support. We
gratefully acknowledge the contributions from the technical staffs of
our home institutions. An allocation of computer time from the Center
for High Performance Computing at the University of Utah is gratefully
acknowledged.  We thank VAISALA for providing NLDN data under their
academic research use policy. Data that supports the conclusions presented in the manuscript are
provided in the figures of the paper and in the additional tables and
figures of the Supporting Information.

%%%%%%%%%%%%%%%%%%%%%%%%%%%%%%%%%%%%%%%%%%%%%%%%%%%%%%%%%%%%%%%%%%%%%%%%%%%%%%%%

%\bibliography{talma_grl}{}

\end{document}